\documentclass[letter]{sig-alternate-2013}

\newfont{\mycrnotice}{ptmr8t at 7pt}
\newfont{\myconfname}{ptmri8t at 7pt}
\let\crnotice\mycrnotice%
\let\confname\myconfname%

\permission{Permission to make digital or hard copies of all or part of this work for personal or classroom use is granted without fee provided that copies are not made or distributed for profit or commercial advantage and that copies bear this notice and the full citation on the first page. Copyrights for components of this work owned by others than the author(s) must be honored. Abstracting with credit is permitted. To copy otherwise, or republish, to post on servers or to redistribute to lists, requires prior specific permission and/or a fee. Request permissions from permissions@acm.org.}
\conferenceinfo{COSN'13,}{October 7--8, 2013, Boston, Massachusetts, USA. \\
{\mycrnotice{Copyright is held by the owner/author(s). Publication rights licensed to ACM.}}}
\copyrightetc{ACM \the\acmcopyr}
\crdata{978-1-4503-2084-9/13/10\ ...\$15.00.\\
http://dx.doi.org/10.1145/2512938.2512956}

\usepackage{flushend}
\usepackage{color}
\usepackage{url}
\usepackage[super]{nth}

\def\sharedaffiliation{
\end{tabular}
\begin{tabular}{c}}

\begin{document}



      

\title{Traveling Trends: Social Butterflies or Frequent Fliers?}
%
%
%
%

\numberofauthors{4} 
%
\author{
%
%
\alignauthor Emilio Ferrara\titlenote{Corresponding author: \texttt{ferrarae@indiana.edu}}\\
\alignauthor Onur Varol\\
\and 
\alignauthor Filippo Menczer\\
\alignauthor Alessandro Flammini\\
\sharedaffiliation
\affaddr{Center for Complex Networks and Systems Research}\\
\affaddr{School of Informatics and Computing, Indiana University, Bloomington, USA}
}
\date{}

\maketitle

\begin{abstract}
Trending topics are the online conversations that grab collective attention on social media.  
They are continually changing and often reflect exogenous events that happen in the real world.
Trends are localized in space and time as they are driven by activity in specific geographic areas that act as sources of traffic and information flow. 
Taken independently, trends and geography have been discussed in recent literature on online social media; although, so far, little has been done to characterize the relation between trends and geography. 
Here we investigate more than eleven thousand topics that trended on Twitter in 63 main US locations during a period of 50 days in 2013. 
This data allows us to study the origins and pathways of trends, how they compete for popularity at the local level to emerge as winners at the country level, and what dynamics underlie their production and consumption in different geographic areas.
We identify two main classes of trending topics: those that surface locally, coinciding with three different geographic clusters (East coast, Midwest and Southwest); and those that emerge globally from several metropolitan areas, coinciding with the major air traffic hubs of the country. 
These hubs act as trendsetters, generating topics that eventually trend at the country level, and driving the conversation across the country. 
This poses an intriguing conjecture, drawing a parallel between the spread of information and diseases: Do trends travel faster by airplane than over the Internet? 
\end{abstract}

\category{}{Human-centered computing}{Collaborative and social computing}[Social media]
\category{}{Information systems}{World Wide Web}[Social networks]
\category{}{Networks}{Network types}[Social media networks]



\section{Introduction} 
\label{sec:introduction}

Social media and online social networks have been widely adopted as proxies to study complex social dynamics, such as the spread of information and opinions  \cite{cha2010measuring,conover2011political,kwak2010twitter,marcus2011twitinfo,wu2011says,xie2011visual} and the emergence of patterns of collective attention \cite{bollen2011twitter,bollen2011modeling,lehmann2012dynamical,weng2012competition}.
Groundbreaking results emerged with the analysis of geographic metadata from social media, allowing for the study of human mobility patterns and social media demographics  \cite{gonzalez2008understanding,kulshrestha2012geographic,mislove2011understanding,noulas2011empirical,scellato2010distance,scellato2011socio,brodersen2012youtube}.

It has been suggested that social media may overcome the spatio-temporal limitations of traditional communication: technologically-mediated systems make it possible to ignore physical and geographic distances \cite{ciulla2012beating,naaman2011hip}.
This, however, does not imply that communication patterns on social media are not affected by physical distances and geographic borders \cite{mocanu2013twitter,onnela2011geographic}. 
In this paper, we explicitly study the role played by geography in driving the main topics of discussion on Twitter: trending hashtags and phrases.

\emph{Trends} represent interesting collective communication phenomena: they are user-generated, continually changing and mostly ungoverned (although orchestrated hijacking attempts have already been observed \cite{budak2011structural,ratkiewicz2011detecting,ratkiewicz2011truthy}). 
So far, trends have been studied as a proxy to detect exogenous real-world events discussed in social media, \cite{aggarwal2012event,becker2011beyond,crooks2012earthquake,sayyadi2009event}, emerging topics, or news of interest for the online community \cite{cataldi2010emerging,leskovec2009meme}.

But trends are also strongly localized in space and time: the temporal and geographic dimensions play a crucial role to determine the success of a trend in terms of spreading and longevity.
We argue that unveiling the spatio-temporal dynamics that drive trending conversations on social media is instrumental to many purposes: from designing successful advertising campaigns, to understanding virality and popularity that characterize some topics.
In this paper we characterize the relation between trends and geography by tracking and analyzing trending topics on Twitter in 63 main locations of the United States and at the country level, for a period of 50 days in 2013. 

\begin{table*}[!th]\centering\tiny
\caption{The list of the 63 trend locations in the United States and the relative total number of trends (thousands) they generated in the period between April, \nth{12} and the end of May 2013.} 
\begin{tabular}{@{}lc | lc | lc | lc | lc | lc | lc@{}}
	\hline 
	Albuquerque	& 6.7	& Cincinnati	& 5.8	&	Greensboro	& 5.8	&	Long Beach	& 6.5	&	New Haven	& 5.6	&	Pittsburgh	& 5.8	&	San Francisco &	5.7\\
	Atlanta	& 5.1	& Cleveland	& 5.4	& Harrisburg	& 6.3	& Los Angeles	& 5.2	& New Orleans	& 6.2	& Portland	& 6.4	& San Jose	&	6.6\\
	Austin	& 5.8	& Colorado Springs	& 6.7	& Honolulu	& 6.5	& Louisville	& 5.9	& New York	& 4.4	& Providence	& 5.9	& Seattle	&	5.9\\
	Baltimore	& 5.8	& Columbus	& 6.0	& Houston	& 5.1	& Memphis	& 6.5	& Norfolk	& 6.0	& Raleigh	& 5.3	& St. Louis	&	5.7\\
	Baton Rouge	& 6.5	& Dallas-Ft. Worth	& 5.3	&	Indianapolis	& 5.9	& Mesa	& 6.6	& Oklahoma City	& 5.8	& Richmond	& 6.2	&	Tallahassee	&	6.3\\
	Birmingham	& 6.1	& Denver	& 6.1	& Jackson	& 6.8	& Miami	& 5.5	& Omaha	& 6.4	& Sacramento	& 5.9	& Tampa	&	5.6\\
	Boston	& 5.0	& Detroit	& 4.8	& Jacksonville	& 6.0	& Milwaukee	& 5.8	& Orlando	& 5.8	& Salt Lake City	& 6.4	& Tucson	&	6.6\\
	Charlotte	& 5.2	& El Paso	& 6.5	& Kansas City	& 5.7	& Minneapolis	& 5.6	& Philadelphia	& 5.1	& San Antonio	& 5.8	& Virginia Beach	&	6.8\\
	Chicago	& 5.2	& Fresno	& 6.6	& Las Vegas	& 5.4	& Nashville	& 6.0	& Phoenix	& 5.9	& San Diego	& 6.2	& Washington	&	4.7\\
	\hline 
\end{tabular}
\label{tab:locations}
\end{table*}

\subsection*{Contributions and outline}

Here we study the distribution, origins, and pathways of trends; the dynamics underlying trend production and consumption in different geographic areas; and the competition among trends to achieve global popularity.
In the remainder of the paper we make the following contributions:

\begin{itemize}
	\item In \S\ref{sub:trend-network} we describe a procedure to build a directed and weighted temporal dependence network to infer the trendsetting and trend-following relationships among locations. 

	\item In \S\ref{sub:distribution} we provide a statistical characterization of trends, describing how they are distributed in space and time.
	
	\item In \S\ref{sub:geography} we highlight a locality effect in the trend sharing patterns: geographically close cities share similar trends. This effect of locality yields the emergence of three geographic clusters in the US, namely East coast, Midwest, and Southwest. But we also uncover a surprising fourth cluster, representing metropolitan areas spread across the country. 

	\item The temporal dependence network is exploited to unveil the pathways that trends follow: in \S\ref{sub:backbone} we reconstruct and reveal the significant backbone of this network that carries the trends across the country.

	\item In \S\ref{sub:trendsetters} we describe two different dynamics that govern popularity of trends at the country level, one for cities in each local geographic area and one for metropolitan areas. We conclude highlighting that the major metropolitan areas shape the country trends significantly more than all other locations in the country.

	\item Finally, in \S\ref{sub:hubs} we propose an interpretation for the trendsetting role of major metropolitan areas, by noting their correspondence with air traffic hubs and conjecturing that trends travel through air passengers, just as infectious diseases. 

\end{itemize}

A more extensive literature review can be found in \S\ref{sec:related-work}.

\section{Experimental setup}  
\label{sec:experiments}

In this section we discuss the methodology we followed to generate a dataset of Twitter trends, and the derived temporal dependence network that allows us to unveil the dynamics of trend production and consumption.

\subsection{Trends dataset}

To build our dataset we monitored in real-time all trends appearing on Twitter for a period of 50 days, starting from April, \nth{12} until the end of May 2013.

The Twitter homepage provides a trends box that contains the top 10 trending hashtags or phrases at any given moment, ranked according to their popularity. 
Oftentimes, a promoted trend is showed in \nth{1} position --- for our analysis we disregarded promoted trends since their popularity is artificially inflated by the advertisement.

Each Twitter user can monitor the trends at the \emph{worldwide}, \emph{country}, or \emph{city} level. 
Twitter has identified 63 locations in the United States, displayed in Figure \ref{fig:cities-clusters}, for which it is possible to follow local trends. 
The full list of locations is reported in Table \ref{tab:locations}.
It is worth noting that some areas are over-represented (for example the East coast and California), while some states (namely, North and South Dakota, Montana, Wyoming, Idaho, and Alaska) are not represented at all.\footnote{This has to do with the fact that the activity on Twitter in those states is very low.}

We deployed a Web crawler to check at regular intervals of 10 minutes the trends of each of these 63 locations and, in addition, those at the country level.
We ended up collecting 11,402 different trends overall: 4,513 hashtags and 6,889 phrases.
Table \ref{tab:locations} also reports how many trends have been observed in each location.

\subsection{Trend pathway backbone network} 
\label{sub:trend-network}

To investigate where trends usually start and how they propagate from city to city, we built a temporal dependence network of the 63 locations of the United States represented in our dataset.

This network is directed and weighted: each node corresponds to one of the 63 cities, and the weight of an arc $e_{ij}$ from node $i$ to node $j$ is increased every time location $i$ exhibits a trend before location $j$.
The weight of arc $e_{ij}$ therefore represents the extent to which city $i$ precedes city $j$ in adopting a trend: the higher the weight, the more often location $i$ sets the trends that location $j$ will later adopt. 

Due to the fact that the adopted dataset contains a large number of trending hashtags and phrases, the network obtained using the procedure described above is fully-connected.
This makes the extraction of relevant connections hard, as each location is connected with all the others and only the weight of the connections vary. 

To ease the analysis we applied to this network an edge filtering technique known as multiscale backbone extraction \cite{serrano2009extracting}.
The goal of this procedure is to retain only those connections that are statistically significant, by removing all edges whose weight does not deviate sufficiently from a null model. The significance level of an edge is determined by a threshold parameter $\alpha$. 
Lowering $\alpha$ progressively removes edges and eventually causes the disruption of the network. 
We tuned $\alpha$ to obtain the backbone network with the minimum number of edges that suffices to maintain all 63 nodes connected ($\alpha=0.3$).
The resulting multiscale backbone of the network is used for the analysis of pathways of trend diffusion, and to investigate trendsetting and trend-following dynamics (see \S\ref{sub:backbone}).

\section{Results} 
\label{sec:results}

The results of our analysis are discussed in this section: after a statistical description of trends, discussing how they are distributed in space and time (\S\ref{sub:distribution}), we explore their geographic dimension, defining what areas of the country share the same type of trends (\S\ref{sub:geography}); then we further investigate the temporal dimension, discussing the pathways trends follow (\S\ref{sub:backbone}), and finally we characterize the trendsetting and trend-following dynamics (\S\ref{sub:trendsetters}).

\begin{figure}[!t]\centering
	\includegraphics[width=\columnwidth]{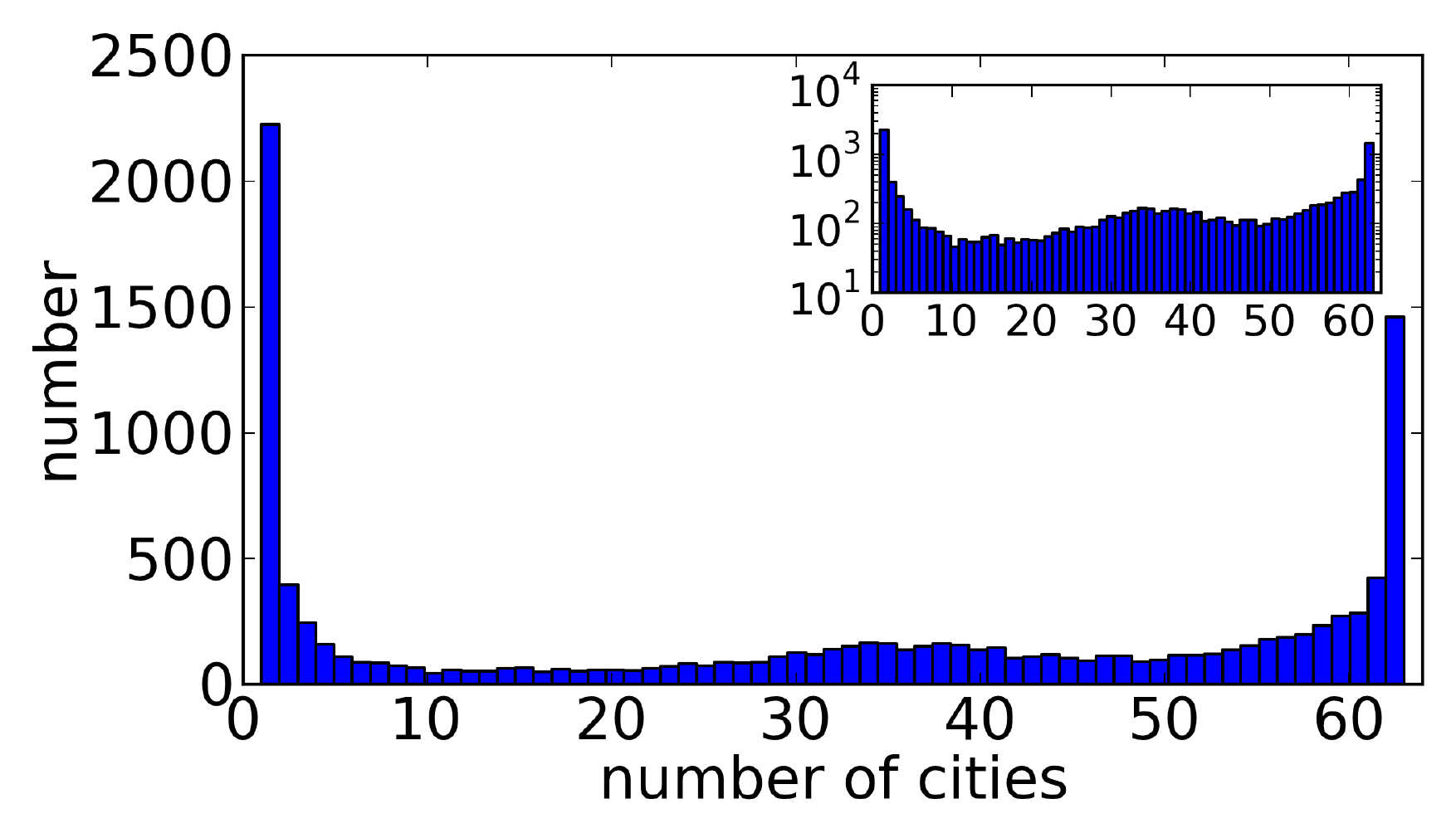}
	\caption{Histogram of the number of trends appearing in different number of places. Inset: y-axis reported in a log-scale.} \label{fig:trends-histogram}
\end{figure}

\begin{figure}[!t]\centering
	\includegraphics[width=\columnwidth]{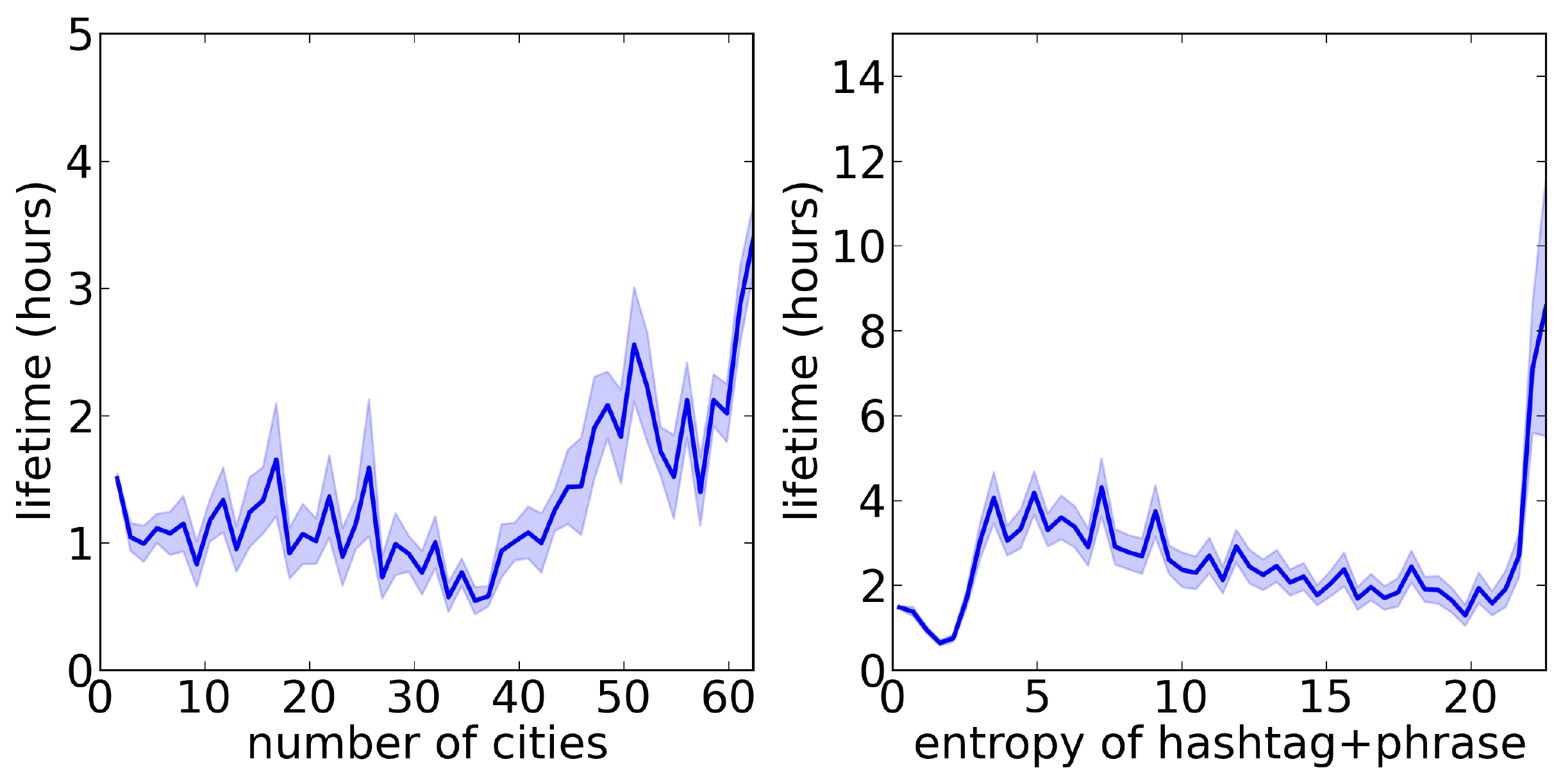}
	\caption{Lifetime of a trend. Left: as function of the number of cities in which a trend has appeared. Right: as function of its entropy. In both plots, the dark blue line is the average across trends while the standard error is depicted in light blue.} \label{fig:life-entropy}
\end{figure}

\begin{figure*}[!th]\centering
	\includegraphics[width=1.8\columnwidth]{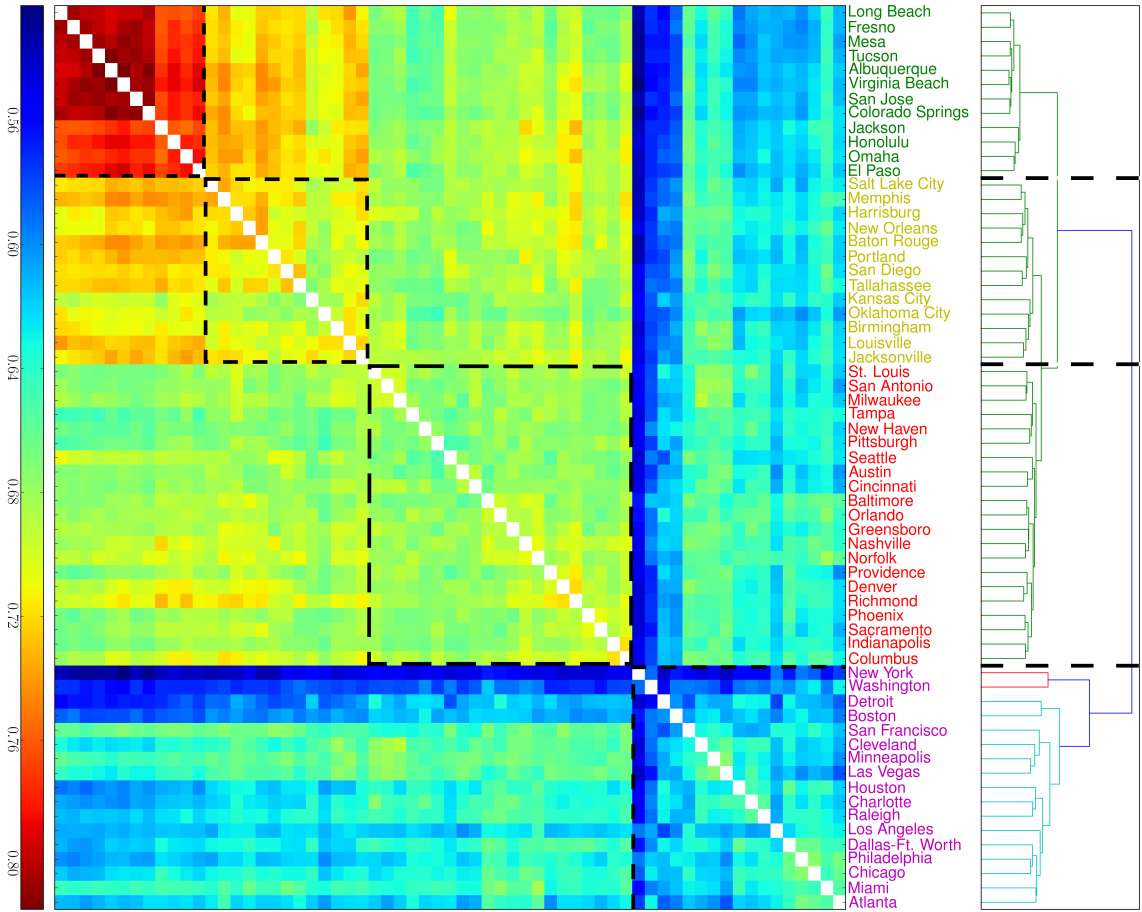}
	\caption{Shared trend similarity and hierarchical clustering of the 63 locations.} 
	\label{fig:content-similarity}
\end{figure*}

\subsection{Spatio-temporal trend analysis} 
\label{sub:distribution}


In our first experiment we aim to give a statistical characterization of trends: in particular, we start investigating in how many different cities trends appear.
In Figure \ref{fig:trends-histogram} we report the number of trends appearing in a given number of distinct locations.
Trends follow a bimodal distribution, typically appearing either in one or few locations, or in all or most of them.
We can identify three behaviors: (i) a large fraction of trends are localized and not sustained enough to spread from their originating place to others; (ii) another comparably large fraction of trends diffuse all over the cities generating a global phenomenon across the country; and (iii) the small remainder diffuse from the originating place to some other places, but fail to achieve global popularity.


The lifetime of trends is broadly distributed: short-lived topics trending for less than 20 minutes amount for more than 68\% of the total, and overall trends shorter than six hours cover more than 95\% of our sample. 
Sporadically some trends happen to live a much longer time, with only 0.3\% surviving for more than a day.

We now focus on the spatio-temporal dimension of trends, aiming to determine how much time each trend spends in one or several locations.
In particular, we calculate the average lifetime of a trend (the average amount of time a given hashtag or phrase is trending somewhere) as a function of the number of cities in which it appears.
Figure \ref{fig:life-entropy} (left panel) reflects the intuition that trends reaching more places live longer.  

Another way to determine the relation between the \emph{geographic spread} of trends and their temporal patterns is to measure their lifetime as a function of \emph{entropy}, defined as 
\begin{equation}
	\mathcal{S}^j = - \sum_{i}{P_i^j \log P_i^j}, \text{\ with \ } P_i^j = \frac{t_i^j}{\sum_{k}{t_k^j}}, 
	\label{eq:entropy}
\end{equation}
where $t_i^j$ is the time topic $j$ has been trending in location $i$.
The entropy is low if the trending topic is concentrated in a few places, and maximal if the topic trends for equal durations of time in all places.
Figure \ref{fig:life-entropy} (right panel) shows that for trends with low entropy (\emph{i.e.}, those concentrated in a single location), the expected lifetime is very short. The lifetime increases significantly (five-fold) for the maximum observed entropy. 
This analysis reveals a key ingredient for global trend popularity: the  trending time of a topic is not only determined by its lifetime in a single location, but also by its geographic spread across many locations.


\begin{figure*}[!th]\centering
	\fbox{\includegraphics[width=2\columnwidth, trim = 50 0 50 0, clip = true]{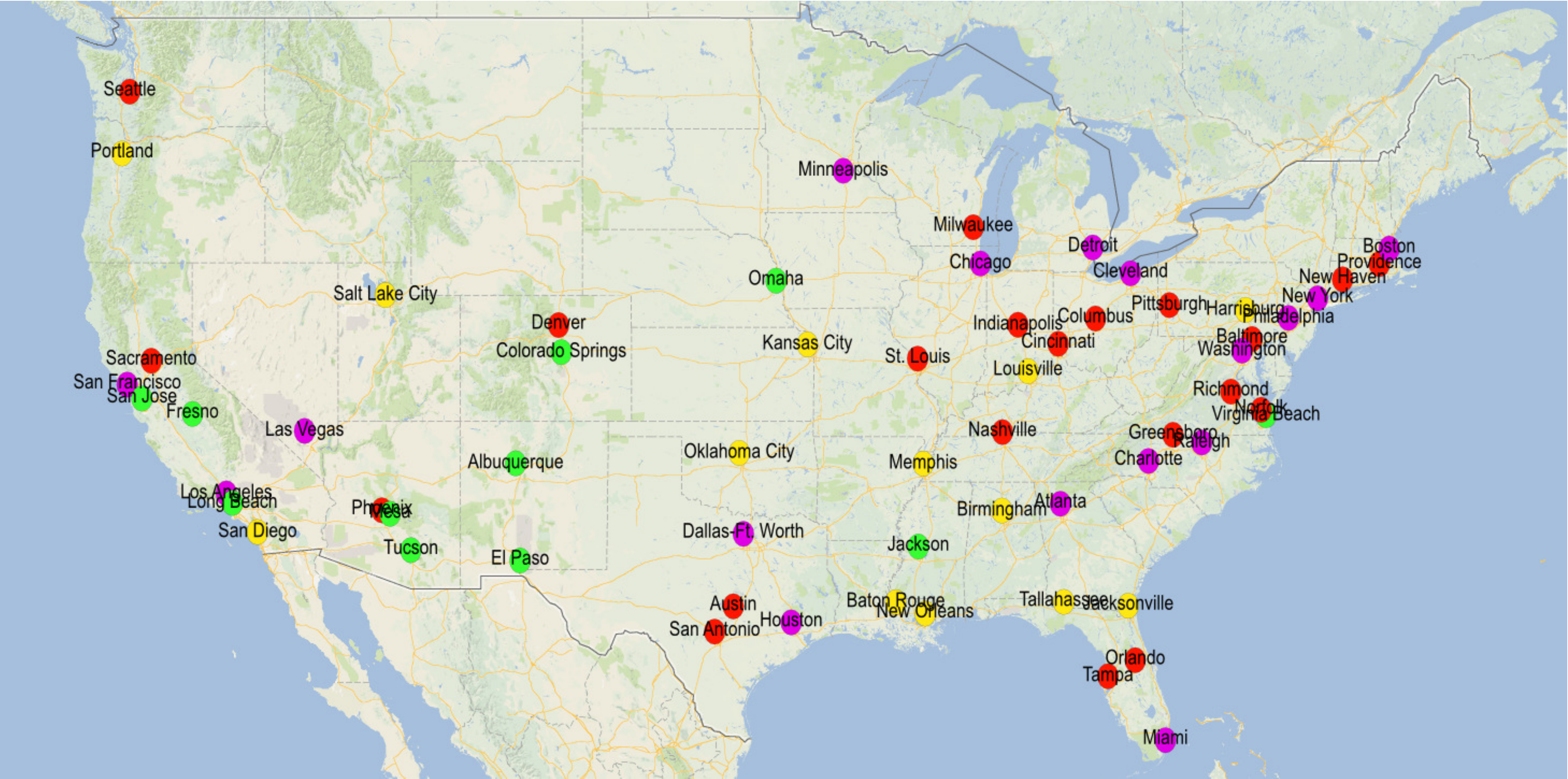}}
	\caption{geographic representation of the 63 locations and respective clusters.}\label{fig:cities-clusters}%
\end{figure*}

\subsection{Geography of trends} 
\label{sub:geography}

Let us examine the geographic patterns of trends, namely whether geographically close cities share more similar trends than cities that are physically far apart.
To determine if this locality effect exists, we first isolate, for each location $i$, the set of trends $T_i$ that appeared in that location. Then, for each pair of locations $i$ and $j$  we compute the pairwise Jaccard similarity 
\begin{equation}
	S_{ij}=\frac{|T_i \bigcap T_j|}{|T_i \bigcup T_j|}.
	\label{eq:jaccard}
\end{equation}
The Jaccard similarity ranges between 0 and 1: the higher the value, the more similar the trends exhibited by two different cities.
These values of similarity are subsequently passed to a hierarchical clustering algorithm after being transformed in distances: $d_{ij} = 1 - S_{ij}$.
This is done to determine whether it is possible to isolate clusters of locations that exhibit similar trends, and, if so, whether these locations are geographically close or spread all over the country.
The result is showed in Figure \ref{fig:content-similarity} and discussed next.


\subsubsection{Locality effects}

Figure \ref{fig:content-similarity} is constituted by two parts: a heat-map representing the pairwise Jaccard similarity among locations, and a dendrogram generated according to an agglomerative hierarchical clustering algorithm using complete linkage. 
Analyzing the dendrogram we can identify three distinct clusters, whose members (reported in different colors: green, yellow and red) share a high internal similarity in the trends exhibited during the observation period. 
This cluster emerges applying a cut to the dendrogram for a distance value of $0.5$.
We can also identify a fourth cluster (in purple, emerging with a dendrogram cut corresponding to a distance value of $0.75$) that exhibits a lower internal similarity and whose members show a low similarity with those of other clusters.
The four clusters are reported in Table \ref{tab:clusters}, and displayed in Figure \ref{fig:cities-clusters}. 

From the figure we observe that the green, yellow and red clusters are somewhat geographically localized, while the purple one is spread more or less all over the country. 
In detail, the green cluster, with the highest internal similarity, roughly corresponds to the Southwest of the country. The yellow cluster follows, representing the Midwest and South. The red cluster, which is less localized, matches many locations in the East coast and Midwest. The purple cluster includes several major metropolitan areas \cite{wiki2011list}; their effect on trendsetting dynamics is discussed in \S\ref{sub:trendsetters} and a conjecture about their role is offered in \S\ref{sub:hubs}. 

\begin{table}[!th]\tiny\centering
\caption{Clusters of cities according to trend similarity. 
}
\begin{tabular}{@{}l | l | l | l@{}}
	\hline 
	\colorbox{green}{\textcolor{black}{Green}}	&
	\colorbox{yellow}{\textcolor{black}{Yellow}}	&
	\colorbox{red}{\textcolor{white}{Red}}	&
	\colorbox{magenta}{\textcolor{white}{Purple}}	\\
	\hline 
	Long Beach		&	Memphis			&	St. Louis	&	Washington\\
	Fresno			&	Salt Lake City	&	San Antonio	&	New York\\
	Mesa				&	Harrisburg		&	Milwaukee	&	Detroit\\
	Tucson			&	New Orleans		&	Tampa		&	Boston\\
	Albuquerque		&	Baton Rouge		&	Pittsburgh	&	San Francisco\\
	Virginia Beach	&	Portland			&	New Haven	&	Cleveland\\
	San Jose			&	Tallahassee		&	Seattle		&	Minneapolis\\
	Colorado Springs&	San Diego		&	Cincinnati	&	Las Vegas\\
	Jackson			&	Kansas City		&	Austin		&	Houston\\
	Honolulu			&	Oklahoma City	&	Orlando		&	Charlotte\\
	El Paso			&	Birmingham		&	Baltimore	&	Raleigh\\
	Omaha			&	Louisville		&	Greensboro	&	Los Angeles\\
					&	Jacksonville		&	Nashville	&	Dallas-Ft. Worth\\
					&					&	Norfolk		&	Chicago\\
					&					&	Providence	&	Philadelphia\\
					&					&	Denver		&	Miami\\
					&					&	Richmond		&	Atlanta\\
					&					&	Phoenix		&	\\
					&					&	Sacramento	&	\\
					&					&	Columbus		&	\\
					&					&	Indianapolis	&	\\
	\hline 
\end{tabular}
	\label{tab:clusters}
\end{table}

\subsubsection{Significance of geographic clustering} 
\label{sub:significance}

\begin{figure}[b!]
	\includegraphics[width=\columnwidth]{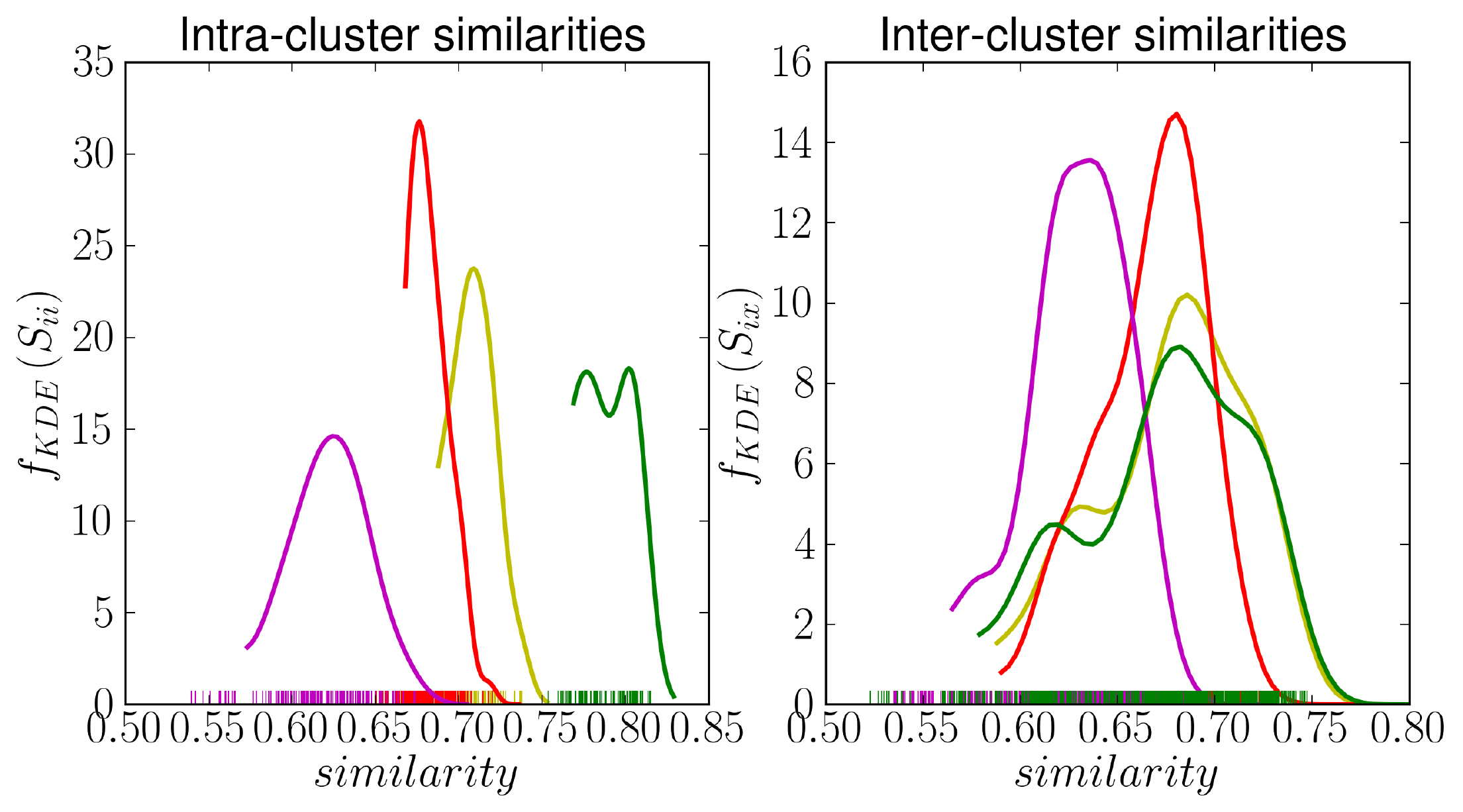}%
	\caption{Kernel Density Estimation of intra- and inter-cluster similarity of the four clusters.}\label{fig:similarity-kde}%
\end{figure}

To determine the statistical significance of the clustering obtained by using the previous method we proceeded as follows: we first computed the distribution of similarity values among all pairs of locations belonging to the same cluster (intra-cluster similarities); then, we did the same for the pairs belonging to different clusters (inter-cluster similarities). 
After that, we applied a kernel smoothing technique known as Kernel Density Estimation \cite{trevor2001elements} to estimate the probability density functions for our similarity distributions, plotted in Figure \ref{fig:similarity-kde} (the distribution of each cluster is represented by its color corresponding to Table \ref{tab:clusters}).

We applied a \emph{t}-test to determine if any given pair of distributions of intra- and inter-cluster similarity might originate from the same distribution, assessing that all distributions (and, therefore, the clusters) are significant at the 99\% confidence level.

We also compared the result of the hierarchical clustering with that of two network clustering algorithms (namely, Infomap \cite{rosvall2008maps} and the \textquoteleft Louvain method\textquoteright~\cite{blondel2008fast}) applied to the trend pathway backbone network (described in \S\ref{sub:trend-network}).
We obtained consistent results in all cases: the only difference was that Seattle was placed in the purple cluster by both network clustering methods.

\begin{figure*}[!th]\centering
	\fbox{\includegraphics[width=2\columnwidth, trim = 50 0 50 0, clip=true]{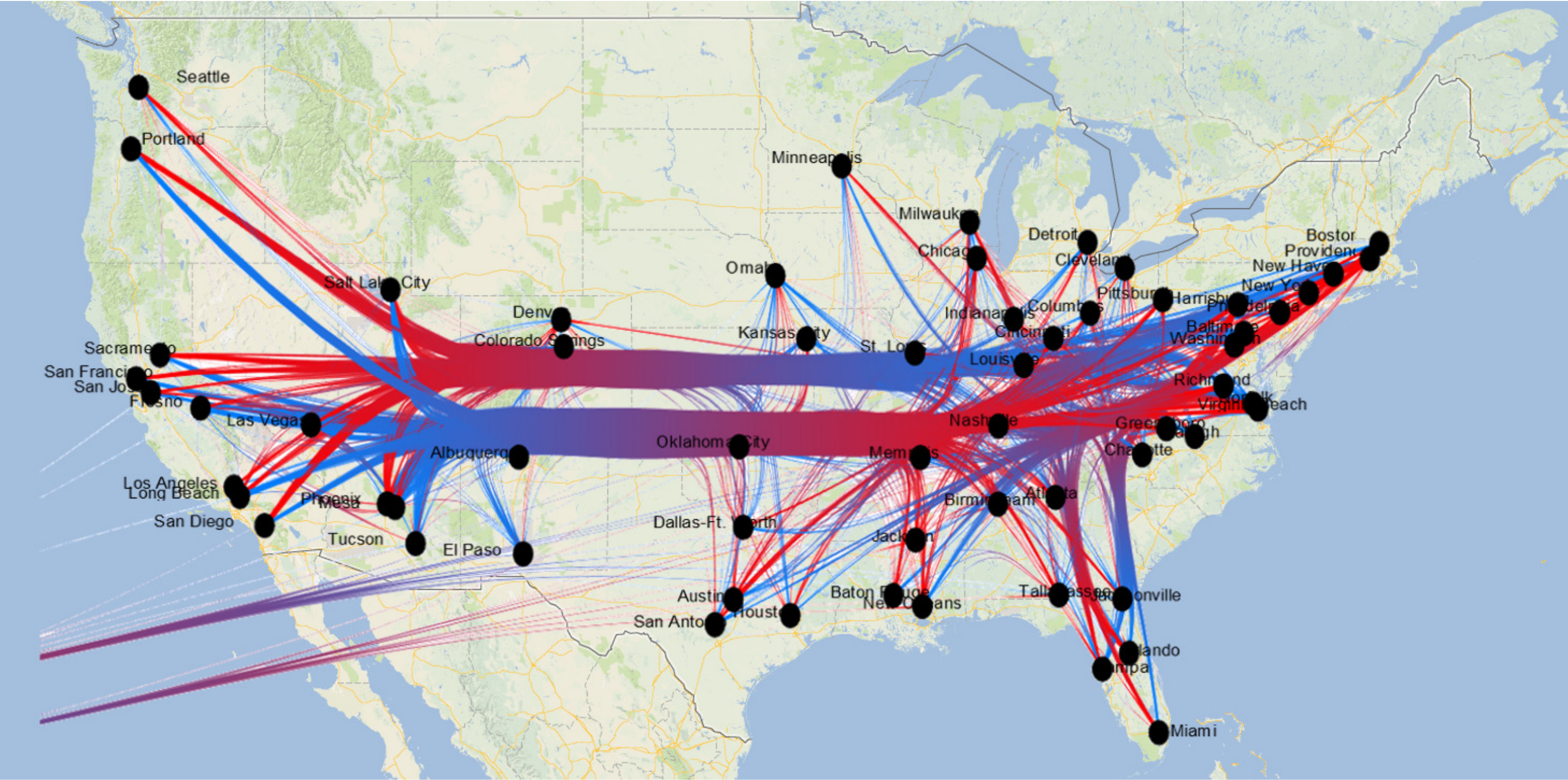}}
	\caption{Trend pathways in Twitter. Trends spread in the direction from blue to red.} \label{fig:information-pathways}
\end{figure*}

\subsection{Trend pathway analysis} 
\label{sub:backbone}

To establish where trends start and what pathways they follow to diffuse in the country, we analyze the multiscale trend pathway backbone network, built as described in \S\ref{sub:trend-network} and represented in Figure \ref{fig:information-pathways} by using a divided edge bundling technique \cite{selassie2011divided}. 
This visualization strategy has been successfully applied to other geographic networks such as the US airport traffic network (\emph{cf.} \cite{selassie2011divided}).
In this node-link representation the edges are bundled taking into account directions and weights. 
The thicker the bundle, the higher the sum of the weights of connections wrapped in the bundle. 
In our case, this yields a network visualization that highlights the pathways followed by trends as they flow across the country.
In this figure the direction of edges represents the information flow: the tails of the bundles (in blue) show where trends start, the heads of the bundles (in red) point to where the trends arrive.
From Figure \ref{fig:information-pathways} we can draw two observations: first, the presence of a massive backbone that carries the trend flow from the East coast to the West coast and vice-versa.
Second, we observe a negligible North-South flow, except for that connecting Florida to the East coast. 
Moreover, the fact that the East-to-West flow is well balanced by the that in the opposite direction suggests that we are not simply observing an artifact of the time-zone effect: the West coast contributes to shaping the country trends to a similar extent that the East coast does.

In the backbone network the cities that often generate trends are those with higher fractions of outgoing edges (that is, those that spread their trends to most of the other cities); henceforth we will call them \emph{sources}.
Vice-versa, we will call \emph{sinks} those cities with higher fraction of incoming edges.
More precisely, since the network we deal with is weighted, we compute the \emph{weighted source-sink ratio} $\omega(n)$ for each node $n$ as  
\begin{equation}
	\omega(n)= \frac{s_{out}(n)}{s_{in}(n)+s_{out}(n)},
	\label{eq:source-sink}
\end{equation}
where $s_{in}(n)$ (resp., $s_{out}(n)$) is the in-strength (resp., out-strength) of that node.
We report in Table \ref{tab:source-sink-ratio} the top 5 sources and the top 5 sinks of the backbone network. 
Four out of the five top sources (all but Cincinnati) also happen to be major metropolitan areas. 
On the other hand, all sinks belong to the Southwest and Midwest parts of the country. 
Los Angeles and New York (among our top sources) have also been reported in the top 5 hashtag producers worldwide in the recent work by Kamath \emph{et al.} \cite{kamath2013spatio}.


\begin{table}[!b]\small\centering
\caption{Left: top 5 sources (\emph{i.e.}, trendsetters). Right: top 5 sinks (\emph{i.e.}, trend-followers).}
\begin{tabular}{@{}lcc|lcc@{}}
	\hline
	Location	&	Rank	&	$\omega(n)$ &	Location	&	Rank	&	$\omega(n)$\\
	\hline
	Los Angeles	&	\nth{1}	&	0.806	&	Oklahoma City		&	\nth{63}	&	0.101\\
	Cincinnati	&	\nth{2}	&	0.736	&	Albuquerque			&	\nth{62}	&	0.109\\
	Washington	&	\nth{3}	& 0.718	&	El Paso					&	\nth{61}	&	0.235\\
	Seattle			&	\nth{4}	& 0.711	&	Omaha						&	\nth{60}	&	0.305\\
	New York		&	\nth{5}	&	0.669	&	Kansas City 		&	\nth{59}	&	0.352\\
	\hline
\end{tabular}
\label{tab:source-sink-ratio}
\end{table}

\subsection{Trendsetters and trend-followers} 
\label{sub:trendsetters}



The source-sink analysis presented above triggered our interest in the dynamics of trend popularity.
In the following we study trendsetting and trend-following patterns, driven by the following question: \emph{Are trending topics that become popular at the country level produced uniformly by all cities, or preferentially by some of them?}

To answer this question we selected from our dataset all those trends that at some point in time became trending at the country level. 
This left us with 1,724 hashtags and 2,768 phrases that achieved the highest popularity in the United States, appearing in the top 10 trending topics at the country level.
We then selected the set of cities that exhibited each of these trends, and divided them in two categories: those cities in which the hashtag or phrase was trending \emph{before} it became trending at the country level, and those cities that adopted it \emph{after} it became trending at the country level. 
This allows us to determine what are the cities that contribute more to shaping the trends at the country level, and what are the cities that are more influenced by these global trends: in other words, we can identify trendsetters and trend-followers.

\begin{figure*}[!th]\centering
	\includegraphics[width=2\columnwidth]{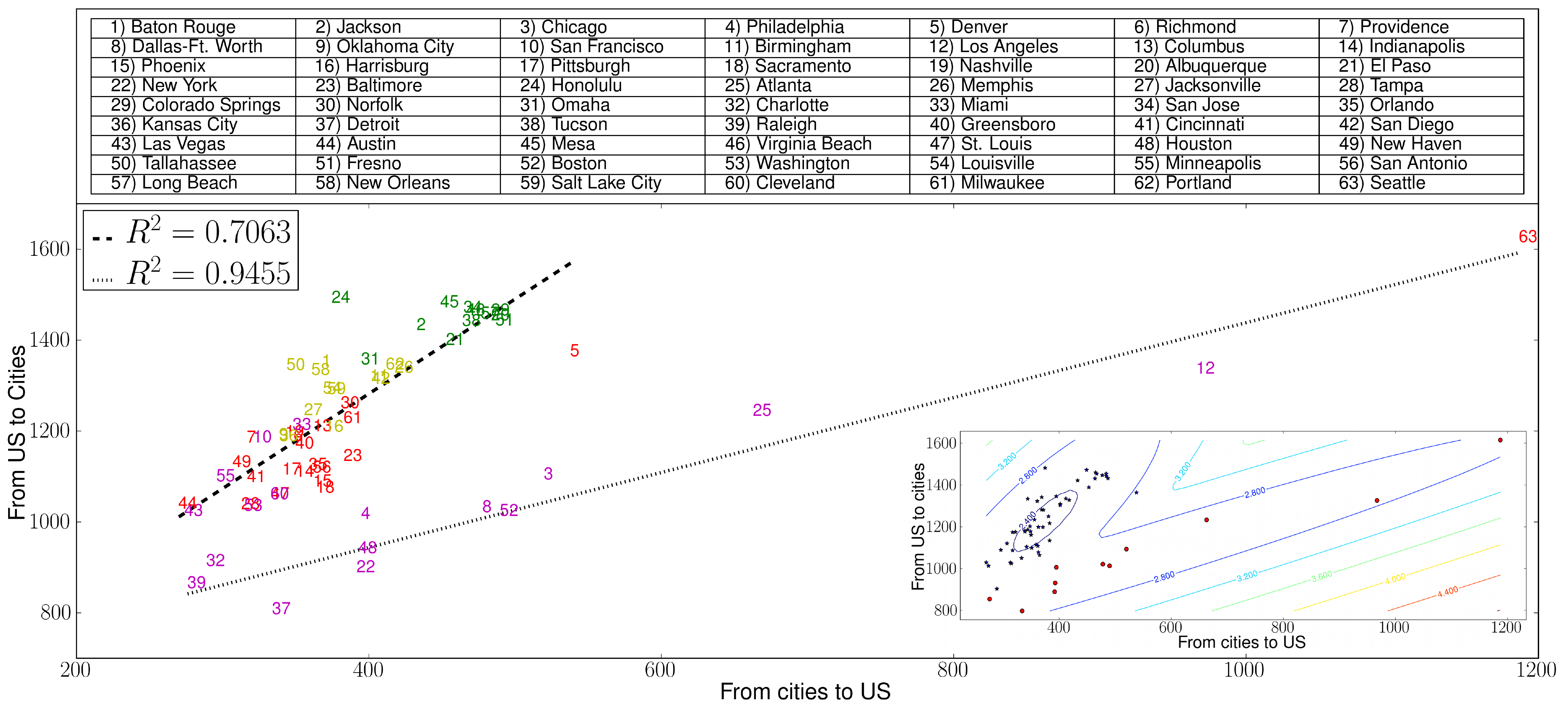}%
	\caption{Trendsetting vs. trend-following cities. The x-axis shows the number of times a topic trending in a particular city later trends at the country level, while the y-axis shows the number of times of the reverse effect. The inset shows a Gaussian Mixture Model highlighting the two different trendsetting dynamics; the contours represent the standard deviations of each Gaussian distribution. In the main plot, two linear regressions are reported with the corresponding coefficient of determination $R^2$. City colors correspond to the cluster assignment in Table \ref{tab:clusters}.} \label{fig:global-cities}
\end{figure*}

Figure \ref{fig:global-cities} shows the result of this analysis for the hashtags. We can immediately identify two different classes of cities: the majority of them (\emph{i.e.}, all those in the upper-left part of the main plot) appear to influence country-level trends roughly to the same extent to which they are influenced by the global trends; a second class of cities seem to have a much stronger trendsetting role toward the country.

To assess if these two classes can be significantly distinguished, we use the Expectation Maximization algorithm to learn an optimal Gaussian Mixture Model (GMM); to determine the appropriate number of components of the mixture we perform a 5-fold cross-validation using Bayesian and Akaike information criteria as quality measures, by varying the number of components from 1 to 10. 
The outcome of the cross-validation determines that the optimal number of components is two, according to both criteria, matching our expectations.

The result of the GMM is showed in the inset of Figure \ref{fig:global-cities}: each point is assigned to one of the two components yielding two different clusters composed respectively of 11 trendsetting cities (red dots) and 52 trend-following cities (blue stars). 
The list of trendsetters includes (in ascending order of impact) Raleigh, Detroit, Philadelphia, Houston, New York, Dallas-Ft. Worth, Boston, Denver, Atlanta, Los Angeles, and Seattle. 
All of them are major metropolitan areas. 

To highlight the existence of these two different dynamics we applied a regression analysis approach by fitting two different linear regressions to the points belonging to the classes of trendsetters (coefficient of determination $R^2 = 0.9455$, p-value $p = 3.9 \cdot 10^{-7}$) and trend-followers ($R^2=0.7063$, $p < 10^{-10}$).
This points out the proportionality that exists between incoming and outgoing trend flows.

We repeated this analysis by making the model even more realistic: for example, we introduced the effect of the time lag, discounting the reward given to those cities that adopt a trend later with respect to the initiators; also, we rewarded only the initiators of each trend, rather than any city that exhibits a given trend before the trending point at the country level. 
Making the scenario more realistic did not affect the outcome: in all cases we obtained comparable results.

\section{Discussion} 
\label{sub:hubs}

The fourth, purple cluster identified in \S\ref{sub:geography} deserves further discussion. 
Differently from the others, this cluster is not geographically well defined (\emph{cf.} Figure \ref{fig:cities-clusters}) --- it contains metropolitan areas spread all over the country. 
Is the effect of city size sufficient to explain why these metropolitan areas are more influential than others, in the sense that they produce more national trends?  
It is not obvious that large populations would lead to more national trends: while a larger city produces more tweets and possibly more topic competing for popularity, the number of trends for each city at a given time is bounded to ten, irrespective of the city size. 
In cities with larger content production, hashtags (or phrases) must appear in more tweets to be listed as a trend, whereas a lower number of tweets is sufficient in cities with smaller content production.
As a result, the effect of sheer volume is discounted by construction in the definition of Twitter trends.

\begin{table}[!tb]\tiny\centering
	\caption{Top 20 cities ranked according to the total volume of flight traffic.}
	\begin{tabular}{@{}lcll@{}}
	\hline
	City	&	Cluster	&	Rank	&	Total traffic\\
	\hline
	New York (JFK, EWR, LGA)	&	\colorbox{magenta}{\textcolor{white}{purple}}	&	\nth{6}, \nth{14}, \nth{20}	&	54,374,758$^*$\\
	Atlanta	(ATL)			&	\colorbox{magenta}{\textcolor{white}{purple}}	&	\nth{1}	&	45,798,809\\
	Chicago (ORD, MDW)		&	\colorbox{magenta}{\textcolor{white}{purple}}	&	\nth{2}, \nth{25}	&	41,603,539$^*$\\
	Miami	(MIA, FLL, PBI)	&	\colorbox{magenta}{\textcolor{white}{purple}}		&\nth{12}, \nth{21}, \nth{54}&	33,228,913$^*$\\
	Dallas-Ft. Worth (DFW, DAL)	&	\colorbox{magenta}{\textcolor{white}{purple}}	&	\nth{4}, \nth{45}	&	31,925,398$^*$\\
	Washington (BWI, IAD, DCA)	&	\colorbox{magenta}{\textcolor{white}{purple}}	&	\nth{22}, \nth{23}, \nth{26}	&31,431,854$^*$\\
	Los Angeles	(LAX)		&	\colorbox{magenta}{\textcolor{white}{purple}}	&	\nth{3}	&	31,326,268\\
	Denver	(DEN)			&	\colorbox{red}{\textcolor{white}{red}}	&	\nth{5}	&	25,799,832\\
	Charlotte/Raleigh	(CLT, RDU)	&	\colorbox{magenta}{\textcolor{white}{purple}}	&	\nth{8}, \nth{37}	&	24,521,523$^*$\\
	Houston	(IAH, HOU)		&	\colorbox{magenta}{\textcolor{white}{purple}}	&	\nth{11}, \nth{32}	&	24,082,666$^*$\\
	San Francisco	(SFO)	&	\colorbox{magenta}{\textcolor{white}{purple}}	&	\nth{7}	&	21,284,224\\
	Las Vegas	(LAS)		&	\colorbox{magenta}{\textcolor{white}{purple}}			&	\nth{9}	&	19,941,173\\
	Phoenix	(PHX)			&	\colorbox{red}{\textcolor{white}{red}}	&	\nth{10}	&	19,556,189\\
	Orlando	(MCO)			&	\colorbox{red}{\textcolor{white}{red}}		&	\nth{13}	&	17,159,425\\
	Seattle	(SEA)			&	\colorbox{red}{\textcolor{white}{red}}	&	\nth{15}	&	16,121,123\\
	Minneapolis	(MSP)		&	\colorbox{magenta}{\textcolor{white}{purple}}	&	\nth{16}	&	15,943,751\\
	Detroit	(DTW)			&	\colorbox{magenta}{\textcolor{white}{purple}}	&	\nth{17}	&	15,599,877\\
	Philadelphia (PHL)		&	\colorbox{magenta}{\textcolor{white}{purple}}	&	\nth{18}	&	14,587,631\\
	Boston (BOS)				&	\colorbox{magenta}{\textcolor{white}{purple}}	&	\nth{19}	&	14,293,675\\
	Salt Lake City (SLC)		&	\colorbox{yellow}{\textcolor{black}{yellow}}	&	\nth{24}	&	9,579,836\\
	\hline
	\multicolumn{4}{l}{($^*$) Sum of the traffic volume of different airports in the same area.}
\end{tabular}
	 \label{tab:top20traffic}
\end{table}

Why, then, do the metropolitan areas in the purple cluster play such a trendsetting role? 
A possible interpretation is offered by noticing the presence in this cluster of some of the major airport hubs of the United States, such as Atlanta, Chicago, and Los Angeles. 
The list of top US airport hubs \cite{wiki2011top20} is shown in Table \ref{tab:top20traffic}, where we aggregated the traffic by metropolitan area. 
Surprisingly, 16 out of the 17 locations that constitute the cluster appear in the top 20 air traffic hubs ---  all of them but Cleveland. 
On the other hand, some cities in the cluster that do not belong in the top 30 metropolitan areas by population (Charlotte, Raleigh, Las Vegas), do appear among the major air traffic hubs. 

The presence of major air traffic hubs among the special class of cities that act as trendsetters suggests an intriguing conjecture, drawing a parallel with the spread of diseases: \emph{Does information travel faster by airplane than over the Internet?} 
In other words, do conversations and trends spread following social interaction dynamics, like \emph{social butterflies} that pass from person to person at the local level, or do they diffuse using traveling people as vectors, similarly to epidemics that take advantage of human mobility \cite{colizza2006role,balcan2009multiscale}?

Further work is needed to explore this conjecture. 
One possibility would be to measure the correlation between trend overlap among pairs of cities and the corresponding air traffic. 

\section{Related work} 
\label{sec:related-work}

Trends or aspects related to geography in socio-technical systems have been studied, directly or indirectly, in many recent studies.
The present work is the first, to the best of our knowledge, that investigates the dynamics tightly binding trends and geography in online social media.

Geographic locations and physical distances have been found to be correlated to friendship behaviors in online social networks \cite{liben2005geographic}, to determine patterns in human mobility networks \cite{brockmann2006scaling,gonzalez2008understanding}, and to affect collaboration schemes in science networks \cite{pan2012world}.

Recent studies took advantage of platforms such as Yelp and Foursquare, which provide customized services to their users based on their physical location (\emph{e.g.}, recommendations of events or places), to study geographic user activity patterns \cite{noulas2011empirical,scellato2011track,scellato2010distance,scellato2011socio}.

Others have used platforms such as Twitter and Facebook, that enrich user profiles with geographic information and accompany user generated content with location-based data, to map users demographics  \cite{kulshrestha2012geographic,mislove2011understanding}.

Onnela \emph{et al.} \cite{onnela2011geographic} noted that, although the probability of observing a tie between two individuals in a social network (in that case, a mobile phone call network) decreases as a power law with physical distance, the geographic spread of social groups quickly increases with the size of the group; even groups of modest dimensions ($\approx$ 30 members) span across hundreds of kilometers, suggesting that, in technologically-mediated social systems, there exist distinctive social dynamics that govern the communication among individuals.

The findings presented in this paper nicely dovetail with Onnela's work, in that we observe the existence of a class of cities, geographically spread across the country, that acts as trendsetters for all other locations. 
On the other hand, we highlight that also a locality effect exists: geographically concentrated areas share similar contents and trends.

The local versus global (``glocal'') nature of communication has been observed before in other types of online conversation \cite{hampton2003neighboring}. In our analysis of the Occupy Wall Street movement on Twitter \cite{conover2013geospatial,conover2013digital}, we noted that geographically localized discussions aim at mobilizing resources (\emph{e.g.}, marshaling financial, material and human capital) while global discourse sets the goals of the movement and develops the narrative frames that reinforce collective purpose.

The influence of the locality effect has been also recently pointed out for innovation adoption on Twitter: Toole \emph{et al.} \cite{toole2012modeling} noted that homophily and physical closeness facilitate the adoption of new technological artifacts, suggesting that the effect of geographic location is critical to describe social dynamics in networked systems.

Geographic factors have also been recently found crucial in the adoption of languages and dialects \cite{mocanu2013twitter}, and in the expression of sentiment \cite{mitchell2013geography,quercia2013dont,quercia2012social} in online social media.
Mocanu \emph{et al.} \cite{mocanu2013twitter} showed how social media data can be used to characterize language geography at different levels of granularity, to highlight patterns such as linguistic homogeneity and linguistic mixture in multilingual regions.
 
Similarly, the study by Mitchell \emph{et al.} \cite{mitchell2013geography} suggests that the adoption of online social media content can be instrumental to describe emotional, demographic and geographic characteristics of users of these socio-technical systems; in particular, they investigated Twitter users active in the US in terms of happiness and individual satisfaction.

Another recent research line related to our work is that of the detection of emerging trends, topics, memes, and events in online social networks and social media \cite{aggarwal2012event,becker2011beyond,cataldi2010emerging,crooks2012earthquake,ferrara2013clustering,leskovec2009meme,mathioudakis2010twittermonitor,sayyadi2009event}.
Naaman \emph{et al.} \cite{naaman2011hip} characterized trends according to different dimensions, such as content, interaction, time-based and social features. 
These features were later used to classify trends, allowing for the identification of exogenous vs. endogenous trends and memes vs. retweet trends. 
In their analysis, the authors did not consider the geographic dimension, that is instead central in this work suggesting that it provides crucial information to characterize trends on online social media.

Finally, social media data can be used to make educated guesses on the outcome of real-word events, such as elections or competitions \cite{digrazia2013more}. 
Ciulla \emph{et al.} \cite{ciulla2012beating} combined trends and geographic information of Twitter data to demonstrate that online social media can be exploited to predict social events in the real-world. 
They collected trending hashtag and phrases related to contestants of the popular TV show \emph{American Idol,} mapping the fan base of each candidate to different geographic regions inside and outside the US, to identify spatial patterns in attention allocation and preferences expressed on the online platform.
These signals were then combined and used to predict voting behaviors of fans, 
achieving good accuracy.

\section{Conclusions} 
\label{sec:conclusions}

In this work we investigated the spatial and geographic dynamics that govern trending topics in Twitter.
We monitored trends from 63 different locations in the United States and, in addition, the trends at the country level, for a period of 50 days. 

We sought to understand how trends are distributed in space and time and how they spread from place to place.
We investigated shared trends among cities, finding that there exists a locality effect whose presence allows for the identification of three broad geographic areas where trends diffuse locally more than globally.
We also identified a fourth cluster of metropolitan areas that counterbalances this locality effect. 
These cities, spread all over the country, 
act as sources of trends for other locations. 
They 
contribute much more than the others to shaping the global trends at the country level.
We finally observed that these metropolitan areas coincide with the major air traffic hubs of the country, suggesting an intriguing conjecture based on a parallel between the spread of information and diseases: Do trends travel faster by airplane than over the Internet? 

Our findings have broad potential applications, that include tailoring online content based on users geographic information, or designing better algorithms for geographic-aware trend prediction. 

As for the future, our analysis opens new research questions that will need further attention. An example is the role of traffic hubs in trend diffusion. More in general,  additional work is needed to understand how to identify locations that can be influential for the spread of a given topic and how to effectively convey the information flow to determine the success of a given commercial campaign.



\subsection*{Acknowledgments}
This work is supported by NSF (grant CCF-1101743), DARPA (grant W911NF-12-1-0037), and the McDonnell Foundation. The funders had no role in study design, data collection and analysis, decision to publish, or preparation of the manuscript.


\bibliographystyle{abbrv}
\bibliography{sigproc}  

%
%

\end{document}